\documentclass{edbk} 
\usepackage[dvips]{graphicx}

\let\footnote\savefootnote

\let\footnotetext\savefootnotetext 

\let\footnoterule\savefootnoterule 

\kluwerbib

\def\ga{\mathrel{\mathchoice {\vcenter{\offinterlineskip\halign{\hfil
$\displaystyle##$\hfil\cr>\cr\sim\cr}}}
{\vcenter{\offinterlineskip\halign{\hfil$\textstyle##$\hfil\cr
>\cr\sim\cr}}}
{\vcenter{\offinterlineskip\halign{\hfil$\scriptstyle##$\hfil\cr
>\cr\sim\cr}}}
{\vcenter{\offinterlineskip\halign{\hfil$\scriptscriptstyle##$\hfil\cr
>\cr\sim\cr}}}}}
\def\la{\mathrel{\mathchoice {\vcenter{\offinterlineskip\halign{\hfil
$\displaystyle##$\hfil\cr<\cr\sim\cr}}}
{\vcenter{\offinterlineskip\halign{\hfil$\textstyle##$\hfil\cr  
<\cr\sim\cr}}}
{\vcenter{\offinterlineskip\halign{\hfil$\scriptstyle##$\hfil\cr
<\cr\sim\cr}}}
{\vcenter{\offinterlineskip\halign{\hfil$\scriptscriptstyle##$\hfil\cr
<\cr\sim\cr}}}}}
\begin{document}



\articletitle[2D Numerical Simulation of Stellar Convection]
             {2D Numerical Simulation of Stellar Convection}

\articlesubtitle{An Overview}
\chaptitlerunninghead{2D Numerical Simulation of Stellar Convection}

\author{Matthias Steffen
\footnote{Institut f\"ur Theoretische Phsyik und Astrophysik, 
Universit\"at Kiel, D-24098 Kiel, Germany}}
\affil{Astrophysikalisches Institut Potsdam, An der Sternwarte 16, D-14482 Potsdam, Germany }
\email{msteffen@aip.de}



\begin{abstract}
The dynamics and thermal structure of the surface layers of
stars with outer convection zones can be studied in some detail
by means of numerical simulations of time-dependent compressible 
convection. In an effort to investigate the properties of 
``stellar granulation'' as a function of spectral type, we have carried out 
elaborate 2-dimensional radiation hydrodynamics calculations of surface 
convection for a variety of stellar parameters.
The main features of these simulations are reviewed, with particular 
reference to standard mixing length models.

\end{abstract}


\section{Introduction}
Convection is a universal feature. Essentially all types of stars have
either a convective core, a convective envelope, or both. Low-mass stars
are fully convective, giants may accommodate several distinct convective 
shells. In the case of the Sun, the energy transport in the inner parts is
entirely due to radiation, while in the outer 28.7\% (in radius) it is 
primarily due to large-scale convective currents. At the surface, the 
solar granulation is the visible imprint of gas flows in the outermost
layers of the convection zone.

The role of stellar convection is far-reaching: Convective energy transport
determines the internal temperature structure of a star and its radius (which 
decreases with increasing convective efficiency), and hence controls the star's 
global properties. Convective regions are chemically completely mixed, 
and overshooting convective flows lead to partial mixing of the adjacent
radiative layers. ``Overshoot'' and similar mixing processes which are not
confined to the convectively unstable regions are thought to
be responsible for the existence of carbon stars, carbon-rich white dwarfs,
and for the destruction of lithium in solar-type stars.
Convective motions and concomitant temperature fluctuations exert a direct 
influence on stellar spectra, causing small but practically relevant changes 
in wavelength position, shape and strength of spectral lines. The stochastic 
convective motions can excite stellar oscillations (like in the Sun) and are
a source of acoustic energy, contributing to the heating of stellar 
chromospheres. Finally, convection and rotation are necessary preconditions 
for the operation of the magnetic dynamo mechanism, and hence for stellar 
activity.

Unfortunately, a closed analytical theory of stellar convection is lacking 
due to the complexity of the underlying hydrodynamical problem.
So far, stellar structure models still rely on a phenomenological 
approximation, the so-called \emph{mixing-length theory} (MLT). 

\section{Hydrodynamical Modeling}

The differential equations governing the physics of stellar convection
are well-known but difficult to solve: the conservation equations of 
hydrodynamics, coupled with the equations of radiative energy transfer.
 While their highly non-linear and 
non-local character precludes an adequate analytical treatment, the 
numerical integration of the system of partial differential equations is 
now feasible. This approach constitutes an increasingly powerful 
method to study in detail the time-dependent properties of a radiating, 
partially ionized fluid under stellar conditions.   
Using modern supercomputers, it is possible to perform 3D
numerical simulations of stellar (surface) convection with realistic 
background physics. But even 2D convection models require 
substantial amounts of computer time, particularly for solving the  
radiative transfer problem.

Just like ``classical'' stellar atmospheres, the hydrodynamical models
are characterized by effective temperature, $T_{\rm eff}$,
surface gravity, $g$, and chemical composition of the plasma.
But in contrast to the mixing-length models, there is no longer any free 
parameter to adjust the efficiency of the convective energy transport. 
Based on first principles, radiation hydrodynamics (RHD) simulations provide 
physically consistent \emph{ab initio} models of stellar convection which
can serve to check the validity of MLT.

\section{State-of-the-Art Results}

In the following, we present some results of elaborate 2-dimensional 
RHD simulations of solar and stellar surface convection.
The models comprise small sections near the stellar surface,
typically extending over 10 pressure scale height in vertical direction,
including the photosphere, the thermal boundary layer near optical depth
$\tau_{\rm Ross}=1$, and parts of the subphotospheric layers. For the hotter 
stars of type A and early F, the surface convection zone(s) are shallow enough
to be entirely fitted into the computational box. For the cooler, solar-like
stars, only the uppermost layers of the deep convection zones can be included
into the model, requiring an open lower boundary.

These simulations are designed to resolve the ``stellar granulation''. 
The effect of the smaller scales, which cannot be resolved numerically, 
is modeled by means of a subgrid scale viscosity (so-called Large Eddy 
Simulation approach). 
Spatial scales larger than the computational box are ignored.
We employ a realistic equation of state, accounting for ionization of
H, HeI, HeII and H$_2$ molecule formation. In order to avoid problematic
approximations (like the diffusion or Eddington approximation), 
we solve the non-local radiative transfer problem along a 
large number ($\approx 10\,000$) of rays in a number of wavelength bands, with 
realistic opacities including the influence of spectral lines. For further
details see Ludwig et al. (\cite{LJS94}) and Freytag et al. (\cite{FLS96}).

\begin{figure}[tb]
\vspace*{-4mm}
\center{\hspace*{-5mm}
\mbox{\includegraphics[width=1.15\textwidth,scale=0.9]{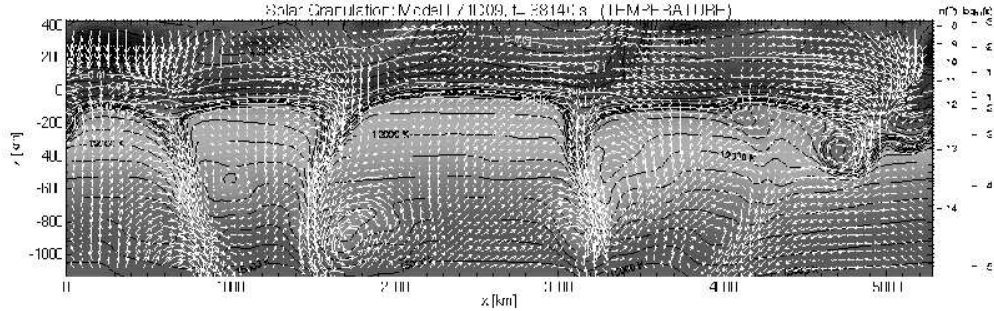}}}
\vspace*{-\baselineskip}
\caption[]{%
\label{L71D09_605}
Representative snapshot from a \mbox{2-dimensional} numerical simulation of 
solar surface convection after 38\,140~s of simulated time. This model was 
computed on a Cartesian grid with 210x106 mesh points (tick marks along 
upper and right side), with periodic lateral boundary conditions 
($L = 5\,250$~km). ``Open'' boundary conditions at the bottom and top of 
the computational domain are designed to minimize artificial distortions of 
the flow. The velocity field is represented by pseudo streamlines, indicating 
the displacement of a test particle over 20~sec (maximum velocity is 10.1~km/s
at this moment); the temperature structure is outlined by temperature 
contours in steps of 500~K. Geometrical height $z=0$ (scale at left) 
corresponds to $\tau_{\rm Ross} \approx 1$; scales at right refer to the 
horizontally averaged gas pressure \mbox{[dyn cm$^{-2}$]} and Rosseland 
optical depth $\tau_{\rm Ross}$.
} 
\vspace{-\baselineskip}
\end{figure}

\subsection{Solar-type surface convection}

\begin{figure}[tb]
\center{\hspace*{-5mm}\scalebox{1.1}[1.0]{
\includegraphics[height=0.75\textheight]{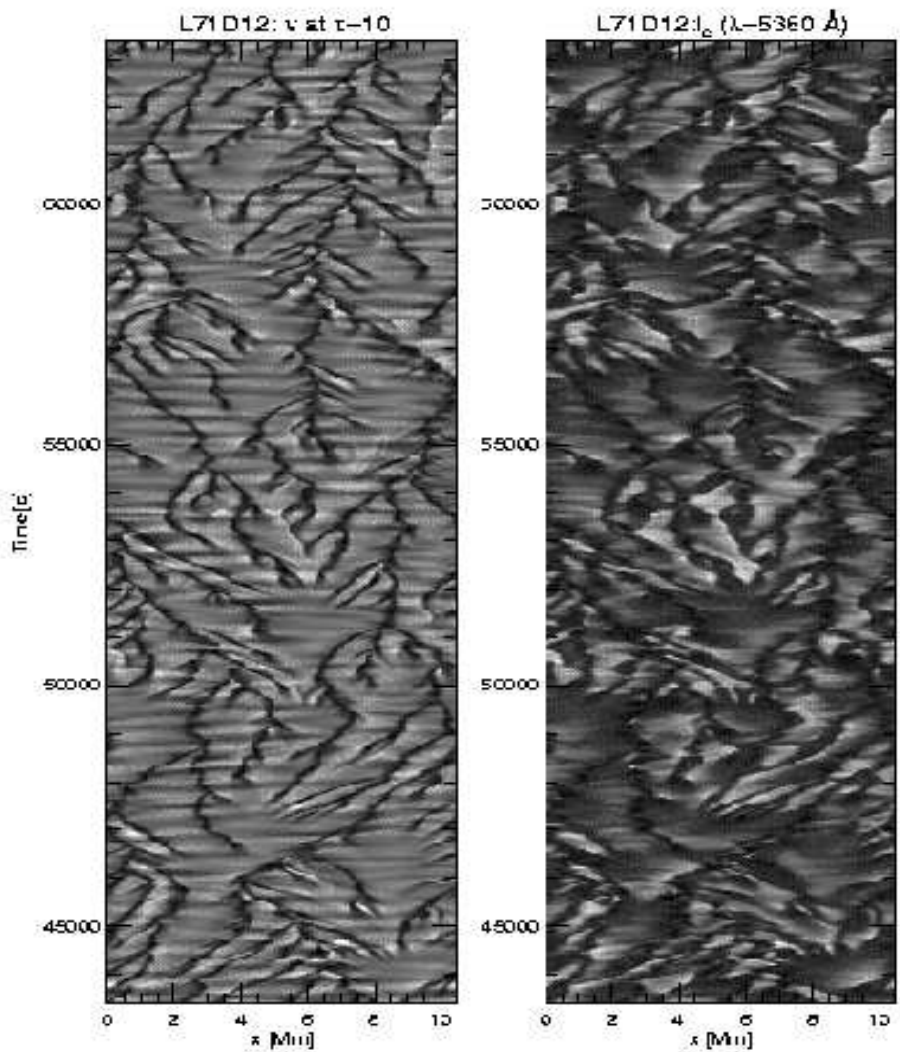}}}
\vspace*{-7mm}
\caption[]{%
\label{L71D12_xt}
Grey-scale plot of vertical velocity at $\tau_{\rm Ross} \approx 10$
(left) and emergent continuum intensity at $\lambda\,5380$~\AA\ (right)
as a function of horizontal position and time. Dark shades indicate
downward velocities and low intensities, respectively. Note that the
signature of superimposed ``5 min oscillations'' is clearly seen in both 
maps. Data were taken from a solar simulation that is similar to the one shown 
in Fig.~\ref{L71D09_605} but has twice the horizontal size (420x106 grid).
} 
\vspace{-\baselineskip}
\end{figure}

A grid of about 60 simulation runs of solar-type surface convection,
covering the range $4\,300$~K $\le T_{\rm eff} \le 7\,100$~K;
 $2.54 \le \log g \le 4.74$ (solar metallicity), is
presently available. In general, the hydrodynamical simulations exhibit 
\emph{unsteady convective flows} and superimposed \emph{oscillations}.

\subsubsection{General properties}
The basic features of the numerical convection models are illustrated in 
Fig.~\ref{L71D09_605}, showing a representative snapshot from a well-relaxed 
simulation of the solar granulation.
In the convectively unstable, subphotospheric layers, fast ($v \la c_s)$
narrow downdrafts (also called plumes or jets) stand out as the most 
prominent feature. They are embedded in broad ascending regions (the granules)
where velocities are significantly lower. We find enormous 
temperature differences between the cool downflows and the hot upflows
($\Delta T_{\rm max} \approx 5\,000$~K at $z \approx -100$~km in the 
solar case). Near optical depth $\tau=1$, efficient radiative surface 
cooling produces a very thin thermal boundary layer over the ascending 
parts of the flow, where the temperature drops sharply with height and 
the gas density exhibits a local inversion. 

Although convectively stable according to the Schwarzschild criterion, 
the photosphere ($\tau \la 1$) is by no means static. Convective flows 
overshooting into the stable layers from below are decelerated here and 
deflected sideways. This can sometimes result in transonic horizontal streams 
which lead to the formation of shocks in the vicinity of strong downdrafts
(see vertical fronts in Fig.~\ref{L71D09_605} near $x=3\,400$~km, 
$z \ga 200$~km).
Oscillations excited by the stochastic convective motions contribute 
to the photospheric velocity field as well. Interestingly, the oscillation
periods lie in the range 150 to 500~s  for the solar simulation, in close
agreements with the observed 5 minute oscillations.
Controlled by the balance between dynamical cooling (due to adiabatic
expansion) and radiative heating (in the spectral lines), the photospheric 
temperature stratification is not at all plane-parallel. At least for the Sun,
the resulting average temperature in the photosphere is slightly cooler than 
in radiative equilibrium.

A typical sequence of events seen in the time-dependent simulations is the 
formation of new downdrafts, their horizontal advection, and finally their 
merging with another downdraft, as illustrated in Fig.~\ref{L71D12_xt}.
We found that merging of downdrafts is a major source of acoustic energy flux. 
Fig.~\ref{L71D12_xt} also demonstrates the strong 
correlation between vertical convective velocity and emergent continuum 
intensity. Downdrafts are clearly seen as narrow dark regions in intensity. 
On the other hand, the correlation is not perfect: in the ascending regions 
there is more structure in the intensity data than in velocity. Small 
``granules'' tend to be brighter than large ones, which often show the 
highest intensity at their edges. Indeed, the power spectra 
of velocity and intensity differ significantly; the p-modes stand out much
more clearly in velocity power.

\subsubsection{Comparison with MLT}
\begin{figure}[!h]
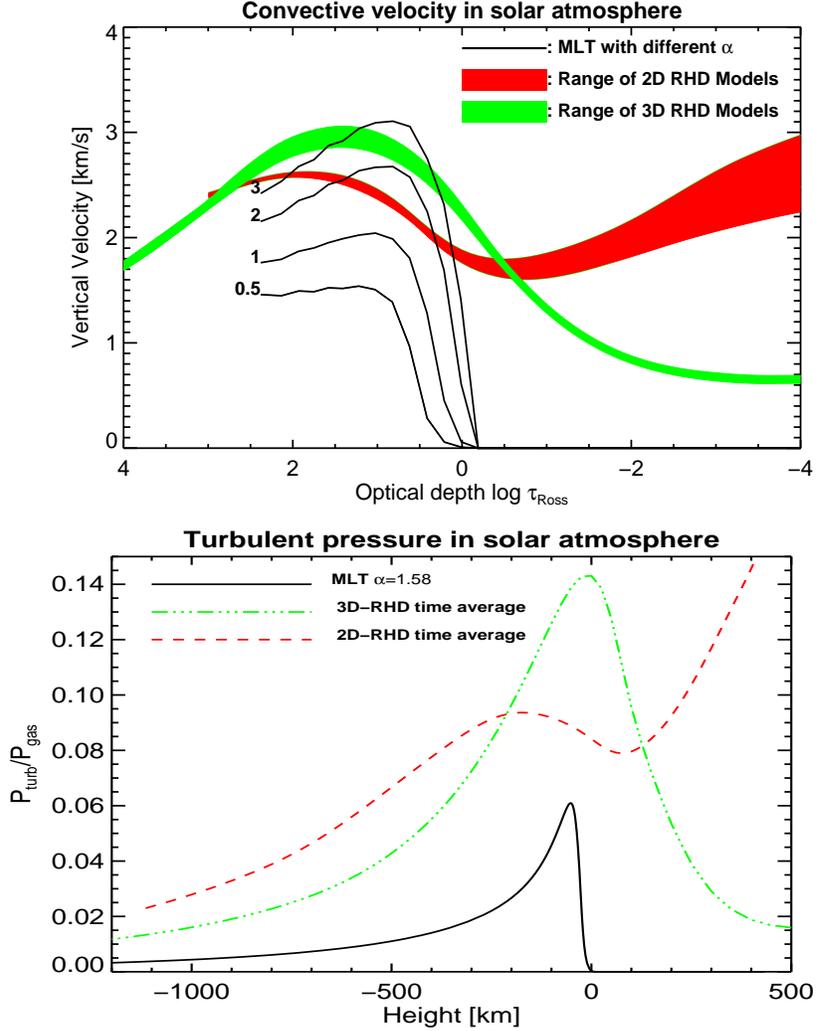

\center{\scalebox{1.0}[1.0]{\includegraphics[bb=20 56 550 400, 
                            width=0.9\textwidth, clip=true]{hk99_mst_F3a.psx}}
        \scalebox{1.0}[0.8]{\includegraphics[bb=75 56 505 400, 
                            width=0.9\textwidth, clip=true]{hk99_mst_F3b.psx}}}
\caption[]{%
\label{MLT_RHD}
{\bfseries Top:} Vertical convective velocity $v$ as a function of optical 
depth $\tau_{\rm Ross}$ as obtained from standard MLT ($\alpha=0.5, 
1, 2, 3$) compared with $v_{\rm rms}$ derived from 2D and 3D numerical 
simulations by averaging over planes of constant $\tau_{\rm Ross}$ and 
over time.
{\bfseries Bottom:} Ratio of turbulent pressure $\rho v^2$ to gas pressure
$P$ as a function of geometrical height as obtained from MLT for $\alpha=1.58$
(adequate for evolution calculations) compared with the
results obtained from 2D and 3D numerical simulations by averaging 
$\rho v^2$ and $P$ over horizontal planes and over time.
Note that a self-consistent MLT solution does not exist. Rather, we obtained
the plotted results \emph{a posteriori} from a standard model computed without 
the effect of turbulent pressure according to the formalism by B\"ohm-Vitense 
(\cite{BV58}), using a non-grey $T(\tau)$ relation that is consistent with the 
numerical simulations. The 3D simulation results
are courtesy of H.-G. Ludwig (see Ludwig \& Nordlund, this volume).
} 
\vspace{-\baselineskip}
\end{figure}

The numerical simulations contain a wealth of information and may be
analyzed in various ways. For comparison with mixing-length theory,
we have averaged the quantities of interest horizontally (over planes 
of constant geometrical height or over surfaces of constant optical depth) 
and over time. Overshoot into the photosphere is substantial according to the 
hydrodynamical models (Fig.~\ref{MLT_RHD}). 
Due to the local nature of MLT, however, it is suppressed in the mixing-length 
models. But even in the unstable layers it is impossible to adjust 
the mixing-length parameter $\alpha$ such as to reproduce the average
2D or 3D velocity field in the framework of MLT. The situation is similar
for the thermal structure $T(\tau)$ (not shown). The difference between the 
2D and the 3D simulation results is probably related to the higher 
acoustic activity of the 2D models (Ludwig \& Nordlund, this volume).

The turbulent pressure $P_{\rm turb}$ amounts to a 
significant fraction of the gas pressure in the solar atmosphere 
(bottom panel of Fig.~\ref{MLT_RHD}) and hence 
affects the geometric scale of the surface layers. Due to the lack of overshoot 
and the uncertainty in the proper choice of $\alpha$, mixing-length models
fail to account for the correct distribution of the turbulent pressure, 
often even assuming $P_{\rm turb}=0$.

\clearpage
\subsubsection {Spectral line formation}
\begin{figure}[bt]
\center{\mbox{\includegraphics[bb=0 60 540 370, width=0.9\textwidth, clip=true]
                               {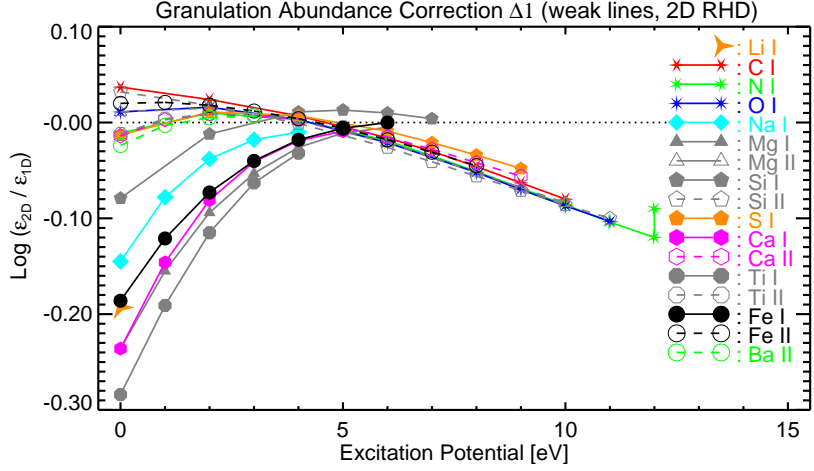}}}
\caption[]{%
\label{delta_ev}
Logarithmic abundance correction to be applied to standard spectroscopic
determinations of solar chemical abundances in order to compensate for the 
effects of photospheric temperature inhomogeneities. These ``granulation
corrections'' depend in a systematic way on the excitation potential of the
spectral line's lower level and on the ionization potential of the atom
or ion being considered. Corrections are valid for weak lines at 
$\lambda 5\,500$~\AA. The NI 12~eV line was also computed for 
$\lambda 10\,000$~\AA\ (upper symbol). In general, 5~eV lines are ``safe''.   
} 
\vspace{-\baselineskip}
\end{figure}

It is not surprising that the photospheric velocity field and temperature 
inhomogeneities due to convective overshoot affect the formation of
spectral lines. We have computed LTE synthetic profiles for a set of weak 
($W_\lambda < 10$~m\AA) fictitious lines ($\lambda = 5500$~\AA) of different
elements and ions, based on an ensemble of representative snapshots from
the 2D solar simulation. For each snapshot, we compared the equivalent width of
the horizontally averaged line profile, 
$W_{\rm 2D}=\langle W_\lambda(\rm RHD)\rangle$, 
with the equivalent width of the same line computed from the averaged (on 
$\tau$-surfaces) RHD model, $W_{\rm 1D}=W_\lambda(\langle\rm RHD\rangle)$.
Averaged over all snapshots, the ratio $W_{\rm 1D}/W_{\rm 2D}$ gives the 
correction factor that has to be applied to standard 1D abundance 
determinations in order to correct for the influence of photospheric 
inhomogeneities. The current results for a variety of different lines
are summarized in Fig.~\ref{delta_ev}. According to this investigation,
lines of neutral minority species originating from the ground state are
most affected. In the most extreme example, the standard analysis based on 
0~eV TiI lines would overestimate the titanium abundance by a factor of 2. 
The effects may be even larger for F-type stars!

\subsubsection{Calibration of Mixing-Length Theory}
As we have seen, the mean dynamical and thermal structure of the
superadiabatic stellar surface layers is systematically different 
in RHD and MLT models, respectively. In this sense, there is no way to 
establish a unique calibration of MLT by means of hydrodynamical simulations. 
However, it is possible to match particular properties by adjusting $\alpha$. 
For stellar evolution, the key quantity is the entropy jump, $\Delta s^\ast$, 
from the surface to the interior. Based on our grid of 
solar-type RHD models, we have calibrated MLT by matching the quantity 
$\Delta s^\ast$, which can be ``measured'' from the simulations. This 
calibration yields the correct adiabat and depth of the convective envelope, 
and hence the appropriate $\alpha^\ast(T_{\rm eff}, \log g)$ for stellar 
evolution calculations. To our knowledge, it is the first theoretical 
prediction of how the mixing-length parameter varies across (parts of) the 
Hertzsprung-Russell diagram (for details see Ludwig, Freytag \& Steffen 
\cite{LFS99}).

For the Sun, the result is
$\alpha^\ast$(2D RHD) $\approx 1.61 \pm 0.05$ and
$\alpha^\ast$(3D RHD) $\approx 1.68 \pm 0.05$, in excellent agreement
with the helioseismic value of $\alpha^\ast$(Helios.) $\approx 1.71 \pm 0.02$.
In the covered range of stellar parameters, we find that $\alpha^\ast$ depends
primarily on $T_{\rm eff}$ and varies systematically between about 1.3 for 
F~dwarfs and 1.8 for K~subgiants. Certainly, the fact that $\alpha^\ast$ varies 
only moderately with stellar type is one of the reasons for the surprising
success of MLT over the past 40 years.

\subsection{Convection in A-type stars}

\begin{figure}[tb]
\center{\hspace*{-5mm}
\mbox{\includegraphics[width=1.15\textwidth]{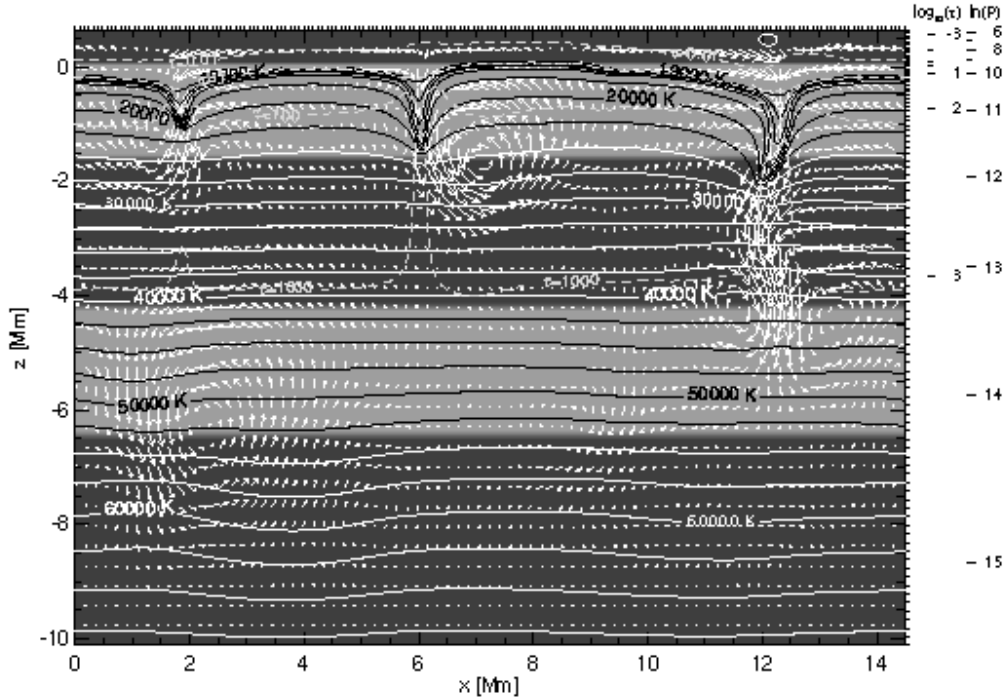}}}
\vspace*{-7mm}
\caption[]{%
\label{at75}
Snapshot from a 2D simulation of convection in an A8 main sequence star 
($T_{\rm eff} = 7\,600$~K, $\log g = 4.4$ [cgs units]) with 2 distinct
convection zones. The unstable regions 
($\mathrm{d}\overline{s}/\mathrm{d}z < 0$, where $\overline{s}$ is the
horizontally averaged specific entropy) are shaded light gray.
The velocity field is represented by pseudo streamlines, indicating 
the displacement of a test particle over 40~sec (maximum velocity is 19.0~km/s
at this moment); the temperature structure is outlined by temperature 
contours in steps of 2500~K. Dashed lines trace levels of constant
$\tau=0.01,1,100,1000$. Located in the lower radiative 
zone, a closed lower boundary is appropriate. Grid size 182x95.
} 
\vspace{-\baselineskip}
\end{figure}

Our grid of hydrodynamical models of A and early F main sequence stars 
comprises about $30$ simulation runs in the range $7\,200$~K $\le T_{\rm eff}
\le 9\,500$~K; $\log g = 4.44$.
These type of stars exhibit two distinct convection zones: 
the one at the surface is driven by the combined first ionization of 
hydrogen and helium, the deeper one is related to the second ionization 
of helium. In the example shown in Fig.~\ref{at75}, the Schwarzschild-unstable 
regions are separated by a stable radiative layer of about two pressure scale 
heights. Nevertheless, the hydrodynamical simulations indicate that the two 
convection zones are effectively connected by vigorous currents and that the 
stable buffer layer is completely mixed, in contrast to MLT predictions. 

A closer inspection of the simulation data showed that the velocity field
due to overshooting convective motions declines exponentially with distance
from the Schwarzschild boundary (in the example of Fig.~\ref{at75}, the 
velocity scale height below the HeII convection zone is $H_{\rm v} 
\approx 0.35\, H_{\rm p}$). We found that mixing due to overshoot can be 
described as a diffusion process, with a diffusion coefficient 
$D \sim v_{\rm rms}^2$. For a slightly hotter star ($T_{\rm eff}=7943$~K, 
$\log g = 4.34$) we have shown that the
convectively mixed mass is underestimated by a factor of 10 if overshoot
is ignored (Freytag, Ludwig \& Steffen \cite{FLS96}).

In stellar structure models, overshoot is usually implemented by 
specifying the distance ($d_{\rm over} = \delta \,H_{\rm p}$) 
by which the convection zone proper is extended due to ``penetration'' into 
the stable layers. This empirical approach ignores the hydrodynamical results
indicating that overshoot is characterized by an exponential
velocity field and (partial) diffusive mixing.

\section{Conclusions}
Radiation hydrodynamics simulations of stellar surface convection
have now reached a level of sophistication far beyond idealized
numerical experiments. They are the key for a better understanding
of the thermal structure and dynamics of stellar convection zones, 
including overshoot and its role for mixing.

The comparison with ``classical'' mixing-length models reveals 
quantitative and qualitative differences.
In the simulations, the dynamics of convection is dominated
by fast, cool, narrow downdrafts, which form coherent structures
extending over many pressure scale heights. A pronounced 
up/down asymmetry is a general feature of the numerical models 
(2D as well as 3D), seen for all types of stars investigated so far. 
Obviously, this result is in stark contrast with the symmetric picture of 
MLT where ``bubbles'' are assumed to travel for about one pressure scale 
height before dissolving.
Since the ``jets'' are driven by radiative cooling at the stellar surface, 
convection is an extremely non-local process. The properties of the surface 
determine the dynamics and structure of the whole convection zone. Certainly,
the assumption of locality is a major problem with MLT.

Hydrodynamical model atmospheres can be used to study the formation of
spectral lines in an inhomogeneous medium. For the Sun, we found that 
``granulation corrections'' for spectroscopic abundance determinations
can amount to -0.3 dex for the most temperature sensitive lines.
These corrections are much larger than the well-known NLTE corrections
that have been investigated so far. Work is under way to check whether
these findings have notable consequences for the currently adopted
chemical composition of the Sun and other stars.

While is seems hopeless to successfully model the structure of the
superadiabatic surface layers within the framework of MLT, it is possible
to calibrate MLT through hydrodynamical simulations for application
to stellar evolution. Based on 2D radiation hydrodynamics, our present
calibration ultimately needs to be verified by 3D simulations. In the meantime,
we are working to extend the calibration to metal-poor stars, down to [M/H]=-2. 

The hydrodynamical simulations for A-type stars (and White Dwarfs)
demonstrate that the velocity field due to overshooting convective 
motions declines exponentially with distance from the Schwarzschild boundary
and leads to diffusive mixing. Although this result is not directly applicable
to strongly adiabatic conditions, we believe that the exponential
depth-dependence of the diffusion coefficient is a general feature of
overshoot and may have important implications for mixing and nucleosynthesis
in stellar interiors.



\begin{acknowledgments}
The author is grateful to H.-G. Ludwig for making available some of his 
3D simulation results and for computing MLT reference models. B. Freytag
computed the A-star model and helped with processing the simulation data.
\end{acknowledgments}



%


\bibliographystyle{apalike}


\end{document}